\documentclass[12pt]{elsarticle}
\usepackage{amssymb,amsfonts,amsmath,mathrsfs,mathtools}
\usepackage{physics}
\usepackage{graphicx,epsfig,subfigure}
\usepackage{bm}
\usepackage[margin=1in,papersize={8.5in,11in}]{geometry}
\usepackage{times}
\usepackage{multirow}
\usepackage{color}
\usepackage{lineno}
\usepackage{capt-of}

\newcommand{\angstrom}{\mbox{\normalfont\AA}}

\title{Improving neural network predictions of material properties with limited data using transfer learning}

\begin{document}
\begin{frontmatter}

\author[ucsc]{Schuyler Krawczuk}
\author[ucsc]{Daniele Venturi \corref{correspondingAuthor}}
\ead{venturi@ucsc.edu}

\address[ucsc]{Department of Applied Mathematics, University of California Santa Cruz, Santa Cruz, CA 95064}

\cortext[correspondingAuthor]{Corresponding author}

\journal{the Journal of Machine Learning for Modeling and Computing}

\begin{abstract}
We develop new transfer learning algorithms to accelerate prediction of material properties from ab initio simulations based on density functional theory (DFT). Transfer learning has been successfully utilized for data-efficient modeling in applications other than materials science, and it allows transferable representations learned from large datasets to be repurposed for learning new tasks even with small datasets. In the context of materials science, this opens the possibility to develop generalizable neural network models that can be repurposed on other materials, without the need of generating a large (computationally expensive) training set of materials properties. The proposed transfer learning algorithms are demonstrated on predicting the Gibbs free energy of light transition metal oxides.
\end{abstract}
\end{frontmatter}

\section{Introduction} 
Determining physical properties of materials using computer 
simulations is commonplace in many areas of 
science, including chemistry, physics and engineering.
This usually relies on ab initio calculations, which are based 
around solving the Schr\"odinger equation \cite{Justin}
for these materials. The  Schr\"odinger equation governs 
the wave function of a many-body quantum system, 
from which information such as its electronic structure 
can be found, and in turn used to determine useful 
properties. Directly solving the Schr\"odinger equation for 
non-trivial systems quickly becomes computationally 
infeasible as the number of atoms increases. 
This led to the rise of methods such as density 
functional theory (DFT) \cite{Parr}, which bypasses 
the need for explicitly solving for the wave 
function to find the electronic structure, 
thus opening the way to perform practical 
computations.
    
The state space of a quantum chemical system 
for even a narrow application of materials can be 
very large, as both content of elements and 
atomic position may be taken into account.
To thoroughly explore this space can take 
significant computational time using DFT, so there 
is a demand to further expedite property prediction 
beyond current state-of-the-art capabilities. 
A leading method for this is machine learning. 
Given sufficient examples of materials, a neural network 
can be trained to represent the mapping between 
atomic structure and material properties such that 
it can generalize well to similar but unseen materials. 
The benefit of this approach is that once trained, a 
neural network can make predictions in a matter 
of milliseconds rather than the minutes or 
hours DFT can take for a single material. 
Several neural network architectures have been 
developed specifically for modelling atomistic systems \cite{behler_generalized_2007,schutt_dtnn_2017,
schutt_schnet_2017}, leading to increasingly accurate representations.
A caveat of current neural network approaches is that it 
can take a large number of examples to learn a sufficiently 
accurate mapping. Although there are large public databases
available with material structures and some of their properties
\cite{jain_materials_2013,curtarolo_aflowliborg_2012}, 
they may not sufficiently cover the target material space, 
leading to the need of generating more training data. 
In these cases, it is desirable to minimize this extra 
cost associated with generating additional data, and 
training accurate neural networks with smaller datasets.
    
This can be achieved by transfer learning, which is a popular 
method used for training neural networks with 
small datasets \cite{pan_survey_2010}. 
Transfer learning reuses representations learned by 
neural networks trained on large datasets. Instead of 
randomly initializing weights for training on a certain task, 
the weights of a neural network trained on a similar 
task are used as a starting point to provide a 
more optimal initialization. Transfer learning has been
very successful in the field of computer vision, 
where large image datasets are publicly available, 
for which pre-trained models are commonly made 
available \cite{Razavian_2014_CVPR_Workshops}. 
These large datasets led to robust machine learning
and generalizable features that are also useful for 
other tasks \cite{pan_survey_2010}. Similarly, in 
material science, large databases of material 
properties have been made publicly available 
\cite{Razavian_2014_CVPR_Workshops}. 
These databases can be leveraged for transfer learning, 
allowing pre-trained neural networks to be 
used for initialization without the creation of 
new data. For instance, in \cite{yamada_predicting_2019} 
transfer learning has been successfully adopted for increasing correlation in material property predictions 
given a very small number of samples. Transfer learning 
has been also used for correcting DFT predictions 
on benchmarks for reaction thermochemistry, 
isomerization, and drug-like molecular torsions
\cite{smith_outsmarting_2018}. 
Along the same line of research as 
\cite{smith_outsmarting_2018}, in this paper 
we develop transfer learning algorithms
to accelerate prediction of material properties 
from ab initio simulations based on density 
functional theory (DFT). 

This paper is organized as follows. In section \ref{sec:DFT} we 
provide a brief review of ab initio calculation of material 
properties using density functional theory. In section
 \ref{sec:NNatomistic} we discuss continuous-filter 
 convolutional neural network representations of atomistic systems (SchNet)\cite{schutt_schnet_2017}, and their training 
using the ADAM algorithm \cite{kingma_adam_2017}. 
In section \ref{sec:SchNetTransfer} we develop transfer 
learning schemes that employ SchNet to improve
training of neural network models with small 
datasets. The accuracy of 
these methods is investigated in 
section \ref{sec:SchNetPrediction} in application  
problems involving transition metal oxides. The main 
findings are summarized in section \ref{sec:conclusions}.

\section{Brief review of density functional theory} 
\label{sec:DFT}
Ab initio calculations for electronic structure rely on solving the time-independent Schr\"odinger equation. The full state of the system is computed by solving an eigenvalue problem of the form 
\begin{equation}
     H\Psi = E\Psi,
\label{eigen}
\end{equation}
where $H$ is the Hamiltonian operator of the system, 
consisting of the sum of kinetic ($T$) and potential 
energies $(V)$, while $E$ represents the energy of 
a specific state (eigenvalue) described by the wave 
function $\Psi$.
In atomic structure calculations, nuclei can be treated as 
a static external potential $V\textsubscript{ext}$, and only 
the electrons are considered in the wave equation \cite{born_quantum_1927}. Hence, the wave function
$\Psi$ in \eqref{eigen} for $N$ electrons in a three-dimensional 
Euclidean space has $3N$ degrees of freedom 
\begin{equation}
       \Psi = \Psi(\bm{r}_1, \bm{r}_2, ..., \bm{r}_N).
\end{equation}
Additionally, each electron has an interaction on each other. As the number of electrons increases, not only does the number of dimensions of the wave function increase, but the number of electron-electron interaction terms in the Hamiltonian increases exponentially \cite{Structure}. In general, the many-body 
Hamiltonian can be written 
as\footnote{In Equation \eqref{vee} $V\textsubscript{e-e}$ 
denotes the electron-electron interaction potential.}
\begin{align}
      H = & T + V\textsubscript{ext} + V\textsubscript{e-e} \label{vee}\\
             = & \sum^N_{i=1} \left[ \frac{-\hbar^2}{2m_e}\nabla^2_{\bm{r}_i} + V\textsubscript{ext}(\bm{r}_i)
            + \sum_{j > i}^N \frac{e^2}{|\bm{r}_{i} - \bm{r}_j|} \right].
    \end{align}
Despite recent advances in 
high-dimensional approximation theory 
\cite{Khoromskaia2018,Han2020,Heyrim2017,Alec2019}, 
solving the eigenvalue problem \eqref{eigen} for a 
non-trivial system involving many atoms quickly 
becomes computationally infeasible. 
    
Density functional theory \cite{Parr} was originally 
proposed to mitigate such dimensionality problem. 
The key idea  relies on expressing material properties 
(ground state) as a functional of the charge 
density $n(\bm{r})$ rather than relying on the 
wave function $\Psi$. The charge density is only 
three-dimensional regardless of the number of 
electrons, leading to a calculations that scale 
with $N$ more efficiently. The fact that ground state 
material properties can be expressed as a functional of the 
charge density was proved in two celebrated theorems 
by Hohenberg and Kohn 
\cite{hohenberg_inhomogeneous_1964}  in 1964. 
For instance, total ground state energy of an atomistic 
system can be written as 
\begin{equation} 
\label{eq:hk_func}
E[n] = F[n] + \int_{\mathbb{R}^3} V\textsubscript{ext}(\bm{r})n(\bm{r}) d\bm{r}, 
\end{equation}
where $F[n]$ is some functional of $n$ \cite{VenturiSpectral,venturi2018numerical} while 
and the second term on the right had side represents 
the interaction of electrons with the external potential 
created from the nuclei. 
The Hohenberg-Kohn theorems showed that the 
functional $F[n]$ exists. However they give 
no guidance on how to find it. 
    
The Kohn-Sham equations were later introduced, which put the Hohenberg-Kohn findings to use and give a practical approach to finding this functional. The Kohn-Sham theory relies on the assumption that a non-interacting system of electrons will have the same electron density as an interacting system of the same structure. Based on this, a Schr\"odinger-like equation can be solved for each individual electron and all resulting wave functions, called the Kohn-Sham orbitals, can be used to calculate the electron density. In this setting, the eigenvalue problem \eqref{eigen} is replaced by a simpler eigenvalue problem of the form  
\begin{equation}
\label{eq:heff}
        H\textsubscript{eff} \psi_i = \epsilon_i \psi_i, 
\end{equation}
where $H\textsubscript{eff}$ is an {\em effective} 
Hamiltonian for this fictitious system. The effective 
Hamiltonian can be written as
\begin{equation}
H\textsubscript{eff} = T[n] + V\textsubscript{Hartree}[n] + V\textsubscript{XC}[n] + \int_{\mathbb{R}^3} V\textsubscript{ext}(\bm{r})n(\bm{r}) d\bm{r}.
\label{Heff}
\end{equation}
This first term at the right hand side represents the kinetic energy 
of ths system. The second term, $V\textsubscript{Hartree}[n]$
is the Hartree potential, which accounts for the repulsion between the electrons
    \begin{equation}
        V\textsubscript{Hartree}[n] =  \frac{1}{2} \int \dfrac{n(\bm{r}')}{|\bm{r}-\bm{r}'|}d\bm{r}'.
    \end{equation}
The third term in \eqref{Heff}, $V\textsubscript{XC}[n]$, 
is known as exchange-correlation potential and it 
approximates more complicated interactions between 
the electrons. The exchange-correlation functional's 
exact form is unknown, but it is also the smallest 
contribution to the total energy. Because of this, it 
can be approximated and still lead to an accurate solution. 
The simplest approximation used for exchange-correlation 
is the local density approximation (LDA) \cite{Parr}. 
As stated by Kohn and Sham, solids can be considered 
close to the limit of the uniform electron gas \cite{hohenberg_inhomogeneous_1964}. 
The local exchange-correlation energy for the 
uniform electron gas is known, written as 
$\epsilon\textsubscript{xc}$. The LDA 
exchange-correlation function can be written as,
\begin{equation}
E\textsubscript{XC}[n] = \int n(\bm{r})\epsilon\textsubscript{xc}(n(\bm{r}))d\bm{r}.
\end{equation}
A more accurate class of functionals that 
build off of LDA is the generalized-gradient 
approximation (GGA). These are exchange-correlation 
functionals that include a term $F\textsubscript{XC}$ 
that is in terms of the gradient of the density, i.e., 
\begin{equation}
E\textsubscript{XC}[n] = \int n(\bm{r})\epsilon\textsubscript{xc}(n(\bm{r}))F\textsubscript{XC}(\nabla n(\bm{r}))d\bm{r}.
\end{equation}
\noindent
The significant improvement in accuracy given by 
GGA functionals led to the wider adoption of 
density functional theory across chemistry and 
material science \cite{Structure}. The exchange-correlation 
potential $ V\textsubscript{XC}$ in \eqref{Heff}  
is the first-order functional derivative \cite{VenturiSpectral}
of $E\textsubscript{XC}[n]$
    \begin{equation}
        V\textsubscript{XC}[n] = \frac{\delta E\textsubscript{XC}[n]}{\delta n}.
    \end{equation}
To solve the eigenvalue problem \eqref{eq:heff}  
the Kohn-Sham orbitals $\psi_i$ are 
usually represented relative to a finite-dimensional 
basis $\{\phi_1,\ldots,\phi_q\}$ as
\begin{equation}
            \psi_{i} = \sum_{j=1}^{q} c_{ij} \phi_j.
\label{exp}
\end{equation}
A substitution of \eqref{exp} into 
\eqref{eq:heff} and subsequent projection onto the basis 
$\{\phi_k\}$ yields a generalized eigenvalue 
problem for $c_{ij}$, which is typically solved with an iterative 
method \cite{blaha_iterative_2010}.

    \begin{figure}[t]
        \centering
        \includegraphics[scale=0.45]{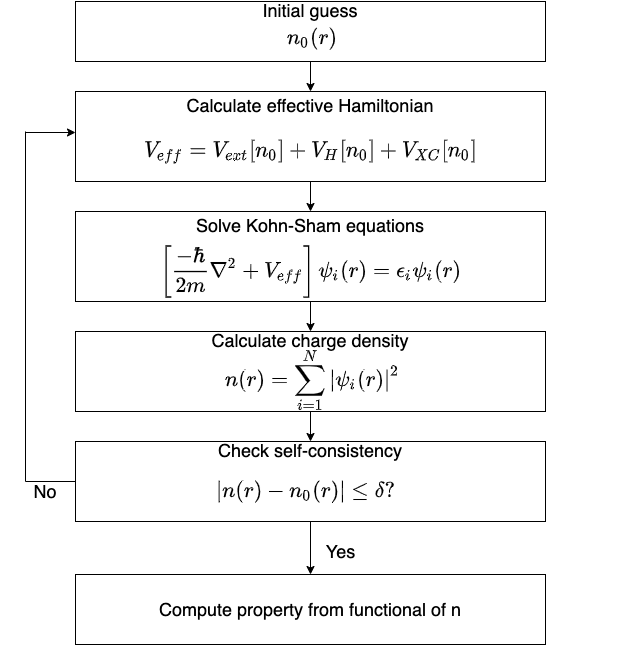}
        \caption{Self-consistent loop for solving the Kohn-Sham equations. First, a guess for the electron density $n(\bm r)$ is created based on atomic structure. Using this $n(\bm r)$, the Kohn-Sham equations are solved and the resulting $n(\bm r)$ is compared to the initial guess. If these are not within a tolerance ($\delta$) of each other, $n$ is updated as a linear combination of the initial and final and another iteration is performed.}
        \label{fig:self_consistent_loop}
    \end{figure}
    \begin{figure}
        \centering
        \hspace*{-1.4cm}
        \includegraphics[scale=0.55]{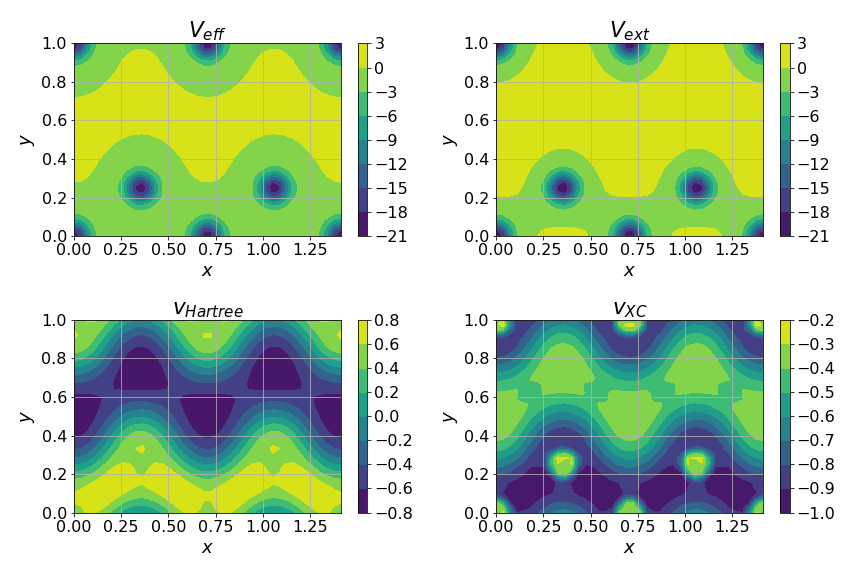}
        \includegraphics[scale=0.55]{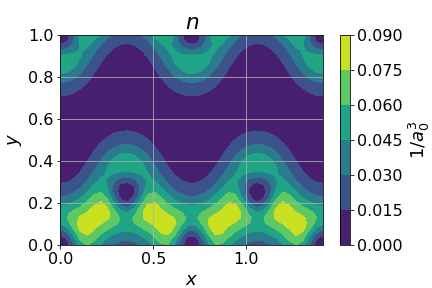}
        \caption{Top: From top left moving clockwise: plot of $V\textsubscript{eff}$, $V\textsubscript{ext}$, $V\textsubscript{XC}$, and $V\textsubscript{Hartree}$ in a two-dimensional slice of a Si$_2$ crystal, along the same plane as the two atoms in the unit cell. Spatial units are in Bohrs ($5.29 \times 10^{-11} $ m) and potential is in Rydberg units ($13.6$ eV). Bottom: The resulting electron density (in electrons per cubic Bohr) for the above potential.}
        \label{fig:my_label}
    \end{figure}
Once the Kohn-Sham orbital $\psi_i$ is computed for each 
electron in the system, the electron density can be 
obtained as
\begin{equation}
      n(\bm{r}) = \sum_{i=1}^{N} |\psi_i (\bm r)|^2. \end{equation}
    This density can then be used to obtain ground state properties. For instance, the total energy function 
    \begin{equation}
        E[n] = T[n] + E\textsubscript{Hartree}[n] + E\textsubscript{XC}[n] + \int_{\mathbb{R}^3} V\textsubscript{ext}(\bm{r})n(\bm{r}) d\bm{r}.
    \end{equation}
The effective Hamiltionian \eqref{Heff} 
requires a charge density to construct it initially, 
meaning an iterative approach needs to be taken to 
finding $n(\bm r)$. If $n$ used to construct the effective Hamiltonian is consistent with the $n$ that results from solving the Kohn-Sham equations, that charge density is consistent for the system. A self-consistent loop, seen in Figure \ref{fig:self_consistent_loop}, is followed 
until consistency is reached. In Figure \ref{fig:my_label} 
we plot the results of a three-dimensional DFT calculation 
for the electron density of a silicon compound.  
    
    \begin{figure}[t]
        \centering
        \includegraphics[scale=0.48]{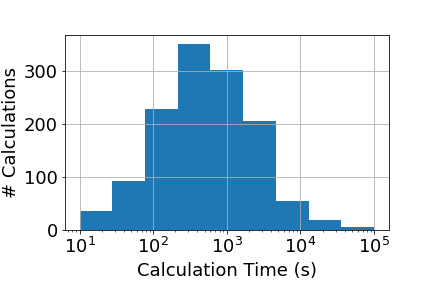}
        \caption{Histogram of DFT calculation times for 1,300 transition metal oxides. DFT simulations were performed on a compute node with 2 Intel 2.1GHz Xeon E5-2620v4 processors (64GB RAM) using 16 cores.}
        \label{fig:dft_timing}
    \end{figure}
    
Computational cost can be a limiting factor in high-throughput calculations with density functional theory. To give a sense of this, 1,300 calculations using the algorithm described above were done for transition metal oxides in this work. These materials had an average of 30 atoms each. Calculations were done on a compute node with 2 Intel 2.1GHz Xeon E5-2620v4 processors (64GB RAM) 
using 16 cores. As shown in Figure 
\ref{fig:dft_timing}, the mean calculation time for 
one run was around 27 minutes, with the quickest taking 5 
seconds and the longest taking 20 hours.

\section{Neural network representation of atomistic systems} 
\label{sec:NNatomistic}
Neural networks are a natural choice for surrogate modeling 
of density functional theory. Their ability to learn 
the nonlinear mapping from an atomic structure to a property has the benefit of not relying on the bias hand-picked features for input, instead solely deriving relationships from the unprocessed data. 
    \begin{figure}[t]
        \centering
        \includegraphics[scale=0.4]{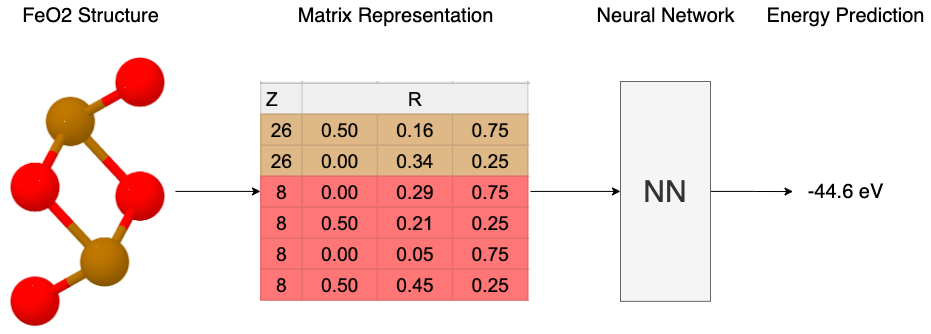}
\caption{Atomistic neural network approach to predicting material properties. The structure of a material can be represented by a matrix consisting of each atom's atomic number (Z) and its Cartesian coordinates (R) with respect to a reference point. This representation will be used as the input to the neural network, which maps it to a target property, such the exchange-correlation potential of the total energy of the compound.}
        \label{fig:atom_nn_input}
    \end{figure}
However, there are some constraints that must be imposed upon the neural network architecture to help enforce the underlying physics.
For instance, different atomistic systems can vary in the number of atoms, while a neural network typically works with a fixed input size. Also, the result should be invariant to the order atoms
are included in the input, as well as the orientation of 
the entire molecule. These characteristics are are not naturally 
accounted for in typical feed-forward neural networks.  
A significant research effort indeed has been recently 
focused in addressing these issues, with specific reference to 
atomistic systems. 

In particular, in \cite{schutt_schnet_2017}
a continuous-filter convolutional neural 
network called SchNet was proposed for 
modeling quantum interactions. 
In our paper we utilize such framework for both 
its nearly state-of-the-art performance on benchmark 
datasets and the SchNetPack toolbox released with it, 
which provides a simple framework for working with 
atomistic systems. SchNet scales well with 
variably-sized inputs by using the same weights for 
each atom in an atomic system, resulting in per-atom 
contributions. Additionally, SchNet employs 
continuous-filter convolutional layers. 
Convolutional layers are a state-of-the-art tools 
for machine learning with spatial data, but typically 
these are discretized, such as pixels of an image. 
Molecular structure does not lie on a grid such as these 
signals. Although it can be discretized, it requires 
choosing a proper interpolation scheme and typically 
a large number of grid points for proper representation 
that can capture subtle positional changes of atoms. 
Continuous-filter convolutional layers are implemented 
in SchNet, getting around this problem by applying a 
convolution element-wise. Given feature representations 
of $n$ objects $\bm{X}^l = (\bm{x}^l_1, ..., \bm{x}_n^l)$ 
at locations $\bm{R}^l = (\bm{r}^l_1, ..., \bm{r}_n^l)$, 
the output of the continuous convolutional layer $l$ 
at position $\bm{r}_i$ is
\begin{equation}
\bm{x}_i^{l+1} = (\bm{X}^l * \bm{W}^l)_i = \sum_{j=1}^{n} \bm{x}_j^l \odot \bm{W}^l(\bm{r}_i - \bm{r}_j), 
\end{equation}
where $*$ denotes the convolution operator, and $\odot$ is the element-wise (Hadamard) product. The filter $\bm{W}^l$ 
weights the distance between the atoms in the system. 
In the continuous-filter convolutional layer, the 
distances $d_{ij} = |\bm{r}_i - \bm{r}_j|$ are expanded 
relative to radial basis functions as
\begin{equation}
        e_k(d_{ij}) = \exp(-\gamma |d_{ij} - \mu_k|^2)
\end{equation}
located at centers $0\angstrom \leq \mu_k \leq 30\angstrom$ 
with $\gamma = 10\angstrom$. Introducing this additional 
nonlinearity causes filter to be less correlated, since the network 
after initialization is close to linear. This speeds up the beginning 
of the training process, which may plateau otherwise 
\cite{schutt_schnet_2017}. 
    \begin{figure}[t]
        \centering
        \includegraphics[scale=0.42]{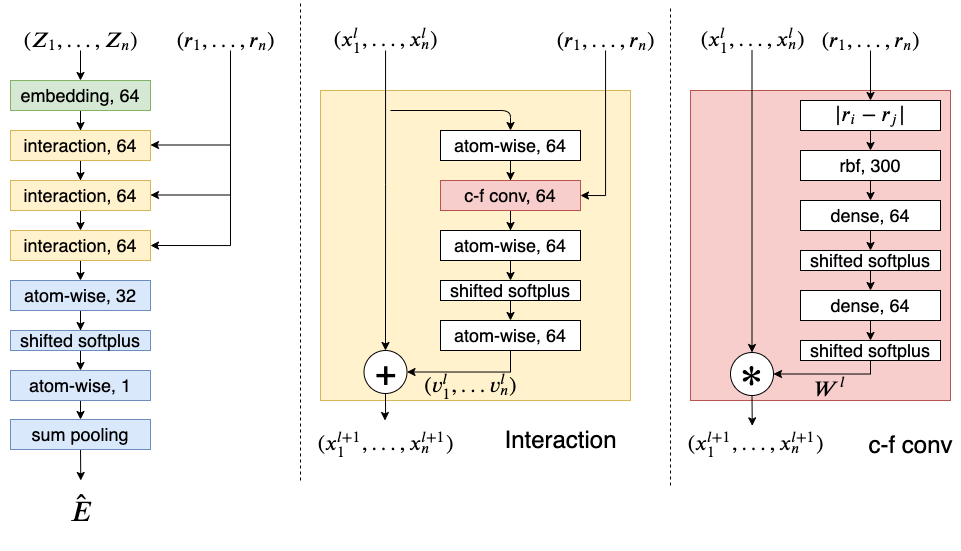}
\caption{SchNet architecture \cite{schutt_schnet_2017} 
using a feature size of 64 and three interaction blocks. 
The interaction block is shown in the middle, and the 
continuous-filter convolutional layer on the right. }
        \label{fig:schnet}
    \end{figure}
In addition to the radial basis functions, each atom 
is represented by an embedding unique to its 
atomic number. This embedding is a vector of a 
predefined length $F$ that is refined through each 
layer $l$ of the network. The feature representations 
for an $n$-atom system for a layer are $X^l = (\bm{x}_1^l, ..., \bm{x}_n^l)$, with $\bm{x}_i^l \in \mathbb{R}^F$. 
Each feature vector is initialized randomly for each 
$Z_i$ such that 
\begin{equation}
        \bm{x}_i^0 = \bm{a}_{Z_i},
\end{equation}
and it is refined during training.

The SchNet architecture consists of the previously described features in blocks called the interaction blocks. Each interaction block refines the feature representations, which are then passed to a final set of atom-wise layers and are pooled to reach the output value. Figure \ref{fig:schnet} shows the full form of this output. 
    Rather than each interaction block being a composition of the previous, as typically done with neural network layers, each uses a residual connection. The features are updated in each layer as 
    \begin{equation}
        \bm{x}_i^{l+1} = \bm{x}_i^{l} + \bm{v}_i^{l}.
    \end{equation}
    This connection helps to prevent overfitting, as it is easier for $\bm{v}_i^{l}$ to become zero in the training process if the next layer is unnecessary by minimizing the residual between $\bm{x}_i^{l+1}$ and $\bm{x}_i^{l}$. Without a residual connection, the next layer would be updated as
    \begin{equation}
        \bm{x}_i^{l+1} = f^l(\bm{x}_i^{l}), 
    \end{equation}
    requiring $f^l$ to learn the identity function, which is a non-trivial task. The activation function used is the shifted softplus, which is defined as
    \begin{equation}
        \text{ssp}(x) = \ln(0.5e^x + 0.5).
    \end{equation}
This function is a smooth approximation of the ReLu. 
The atom-wise layer in the SchNet architecture 
shown in Figure \ref{fig:schnet} applies an affine 
transformation to the features of each atom separately. This layer 
shares the same weights throughout every atom, giving the output 
\begin{equation}
        \bm{x}_n^{l+1} = \bm{W}^l \bm{x}^l_i + \bm{b}^l
\end{equation}
for the atom $i$. The sharing of weights across all 
atoms allows for the network to scale with the size
of the system properly.

\subsection{Training the SchNet architecture}

\begin{figure}
        \centering
        \includegraphics[scale=0.6]{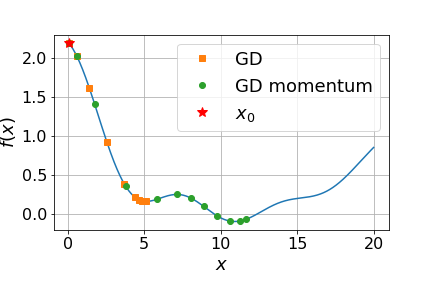} 
        \caption{One-dimensional function minimization using gradient descent with and without momentum. Gradient descent with momentum gradually increases the step size during the descent, allowing it to escape the local minimum at $x=5$, which gradient descent converges to.}
        \label{fig:momentum}
\end{figure}

To train SchNet in a supervised learning setting we 
minimize the mean square error between the predicted 
property and its observed value using the adaptive moment 
(ADAM) algorithm. ADAM is a variation of the 
stochastic gradient descent. In classical stochastic 
gradient descent the weights of the neural 
network $\bm{\beta}$ are updated based on 
the gradient of the cost function $E$ as 
\begin{equation}
    \bm{\beta}_{k+1} = \bm{\beta}_{k} - \alpha \nabla E,
\end{equation}
for step $k$ and learning rate $\alpha$. It is possible to 
improve the convergence rate of stochastic gradient 
descent by multiplying the learning rate by a factor 
of the previous iteration's step. In this setting, each iteration, 
$\bm{\beta}$ is updated as
\begin{equation}
        \bm{\beta}_{k+1} = \bm{\beta}_{k} + \bm{v}, 
        \label{Beta}
\end{equation}
where
\begin{align}
\bm{v} = \eta \bm{v} - \alpha \bm{g}, \qquad 
        \bm{g} = \nabla E. 
 \label{iter}
\end{align}
In \eqref{iter} $\eta \in [0,1)$ is a predetermined parameter \cite{Goodfellow-et-al-2016}, and $\bm{v}$ must be initialized. 
The algorithm \eqref{Beta}-\eqref{iter} is also known as 
gradient descent with ``momentum'',  in a 
physical analogy for the velocity of a ball rolling down 
a hill. In fact, as the balls rolls 
down the hill, much like the optimization descends 
toward a minimum, the ball will gain speed. 
The larger $\eta$ is, the more the previous iteration 
will affect the next one. This momentum also helps 
to escape local minima, as seen in Figure \ref{fig:momentum}.
    
The ADAM algorithm updates individual learning rates based on previous learning rates in an exponential moving average \cite{kingma_adam_2017}. Each parameter is 
updated similarly to \eqref{Beta}-\eqref{iter}. 
Specifically, in ADAM we have that \eqref{iter} is replaced by   
    \begin{equation}
        \bm{v} = \eta \bm{v} - \frac{\alpha}{\sqrt{\bm{r}}} \odot \bm{g}, 
    \end{equation}
where $\bm{r}$ (accumulation of the gradient), 
is updated as
\begin{equation}
    \bm{r} = \rho \bm{r} + (1-\rho) \bm{g} \odot \bm{g}.
\end{equation}
Although ADAM exhibits faster convergence to a minimum, it has been shown to not reach an optimal solution as well as regular stochastic gradient descent, which is more likely to reach a value close to the global minimum \cite{wilson_marginal_2017,reddi_convergence_2019}. 
Similarly to the training scheme used with SchNetPack's 
released models, an initial learning rate $\alpha$ 
of $10^{-4}$ is used in this paper. For each training 
plateau, where the training loss does not decrease 
for 25 iterations, the learning rate is reduced by a 
factor of $0.8$ until a minimum 
learning rate of $10^{-6}$ is reached. 

\section{Transfer learning with SchNet} 
\label{sec:SchNetTransfer}
In this section we develop a transfer learning scheme 
in conjunction with SchNet to improve the ability of models to 
generalize when trained with small training sets. 
Transfer learning reuses representations learned in 
training with a source task as a starting point for 
training on a different but similar target task. In this case, 
a SchNet model is trained with a large source dataset, 
and the optimized weights are reused for 
initializing training with the target dataset. 
    
\subsection{Transition metal oxide database}

In this paper, transition metal oxides are considered 
for property prediction. Transition metal oxides are 
used in practical applications for solar energy 
conversion. However, poor conductivity and 
electron-hole separation limits their carrier conductivity. 
It has been shown that appropriate doping (adding of impurity) 
of these materials may improve their utility. An important 
property to be found in these doped transition metal 
oxides is a low defect formation energy. The defect 
formation energy is the difference between the total 
free energy of the pure transition metal oxide and that 
of the impure, doped transition metal oxide. Although 
there are other properties of importance, this work focuses 
on the learning of the mapping between transition 
metal oxides and their free energy. 
    
The target dataset we first considered is composed of 
517 transition metal oxides of Iron, Titanium, 
and Vanadium, along with the corresponding free 
energies of the compounds. 
These materials each have between 2 and 110 atoms, 
with a mean of 12. An extension of this dataset was 
then created for further testing of the transfer learning 
methods, introducing an additional 146 transition 
metal oxides consisting of Chromium and Manganese.
The source dataset is a subset of the Materials Project 
database \cite{jain_materials_2013}. 
This subset consisted of materials made up of 87 
elements, including those of the target dataset, 
Ti, Fe, V, and O. Materials of the same unit cell 
formula as the transition metal oxide dataset were 
excluded for the sake of preventing overlap between 
the two datasets. This means that materials with 
the same composition as one in the transition metal 
oxide set, even if they had a unique geometry, 
were not included in this dataset. This dataset 
includes 50,000 materials used for training, and 10,000 
for validation. A smaller subset of this dataset was 
created, also excluding Mn and Cr for the purpose 
of transfer learning with the extended 
transition metal oxide dataset.
    
\subsection{Transfer learning schemes}
    
\begin{figure}
        \centering
        \includegraphics[scale=0.6]{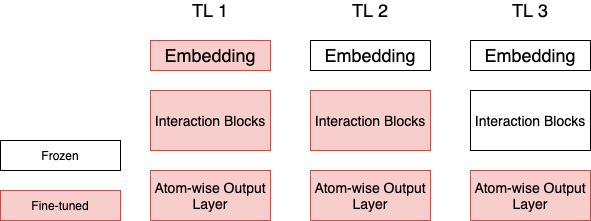}
        \caption{Visualization of the three transfer learning schemes we implemented. The embedding layer, the interaction blocks, and the output layers are either frozen or fine-tuned.}
        \label{fig:tl_schemes}
\end{figure}

We investigated three transfer learning schemes. 
These either chose to freeze weights, not updating 
them in the new training, or to fine-tune them 
by training with the new dataset. The schemes 
used are as follows (see Figure \ref{fig:tl_schemes}) 
\begin{enumerate}
        \item All weights are fine-tuned (TL1);
        \item The embedding layer is frozen and the rest of the weights are fine-tuned (TL2);
        \item Only the output layers are fine-tuned and all other weights are frozen (TL3).
    \end{enumerate}
To obtain a performance benchmark for the 
target dataset, SchNet is trained with a random 
initialization on a training set of 400 transition 
metal oxides. Two different model architectures 
and regularization schemes were used.
    \begin{table}
        \centering
        \begin{tabular}{c|c c c c} 
             $N$ & arch 1 & arch 1 + L2 & arch 2 & arch 2 + L2 \\ [0.5ex] 
             \hline
             400 & 0.563 & $\bm{0.551}$ & 1.175 & 1.005 
         \end{tabular}
        \caption{Validation set mean absolute error in eV (electronvolts) of SchNet models trained on the full transition metal oxide training set. Arch 1 is the larger SchNetPack 
architecture which consists of six interaction blocks, 
128 length embedding vectors and 128-filter 
convolutional layers. Arch 2 is the smaller architecture, 
with four interaction blocks, with embeddings of length 
30 and convolutional layers with 30 filters. 
L2 signifies the inclusion of L2 regularization with a coefficient of $10^{-3}$ in the loss function during training.}
\label{tab:schnet_tmo_results}
\end{table}
Both model architectures used for training on a baseline 
dataset are based on SchNetPack 
\cite{schutt_schnetpack_2019}, with and 
without L2 regularization on all weights. 
The first one consists of six interaction blocks, 
128 length embedding vectors and 128-filter 
convolutional layers.
Since the transition metal oxide dataset is considerably 
smaller than the baseline which consisted of 
130,000 materials, a smaller architecture was 
also trained with and without regularization in 
an attempt to prevent potential overfitting. 
This architecture had four interaction blocks, with embeddings of length 30 and convolutional layers with 30 filters. The results of the validation set evaluated by these models are reported in Table \ref{tab:schnet_tmo_results}. The model with the best mean absolute error was the original architecture with an L2 regularization coefficient of $10^{-3}$. Out of the validation set, only 47 of the 117 predictions were within chemical accuracy. This best-performing architecture and regularization is used in comparison with transfer learning models.

\section{Prediction with SchNet transfer learning}    
\label{sec:SchNetPrediction}
Transfer learning methods are compared to direct 
training methods on the same dataset for varying 
training set sizes. The target data was split into a 
training set of 400 and a validation set of 117 that 
remains the same for all evaluations. Training data 
sizes of 400, 200, and 100 are used. For the 200-length 
dataset training, the 400 are split into two sets, and a 
separate neural network is trained with each. The 
same is done with the 100-length training set, 
where four neural networks are trained. The mean 
absolute error on the validation set of each of 
the same-length dataset networks is then averaged 
to get the given results. This is done to ensure 
consistent results, since the dataset is small and all 
portions may not be entirely representative of each other. 
\begin{figure}[t]
       \centering
\includegraphics[scale=0.55]{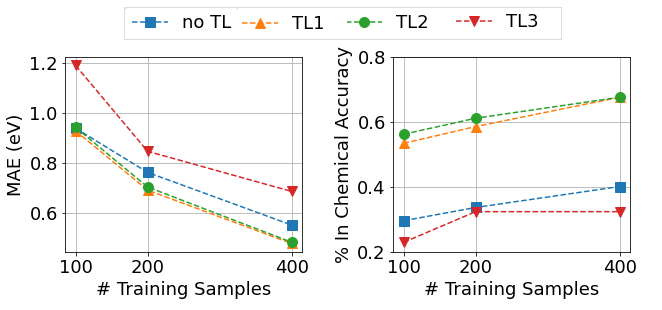}
\caption{Results of the three best transfer learning methods 
compared to best direct training method without transfer 
learning with the original dataset. Left: Mean absolute error 
(MAE) of the validation set evaluation for the models 
trained on each of the training set sizes. Right: Percent 
of evaluations within chemical accuracy of the value 
computed with DFT.}
\label{fig:full_results}
\end{figure}

\begin{table}[t]
        \centering
        \begin{tabular}{c|c c c c} 
             $N$ & No TL & TL1 & TL2 & TL3 \\ [0.5ex] 
             \hline
             100 & 0.939        & $\bm{0.929}$      & 0.943    & 1.191 \\ 
             200 & 0.762        & $\bm{0.691}$      & 0.703    & 0.847 \\
             400 & 0.551        & $\bm{0.477}$      & 0.482    & 0.686 
         \end{tabular}
        \caption{Validation set mean absolute error in eV (electronvolts) of each of the transfer learning schemes along with the best performing direct training method across the different-sized splits of the dataset.  The bold numbers are the best result for the row.}
        \label{tab:schnet_tl_results}
    \end{table}
     
Here, transfer learning methods TL1 and TL2 proved 
to be the most effective, achieving lower mean absolute 
error on the validation set than the best direct training 
model for the 400 and 200 size training sets, and similar 
error for the 100 size training sets.  The third transfer learning 
scheme performed worse than the direct training. 
The mean absolute errors are reported in Table 
\ref{tab:schnet_tl_results}. A significant difference was seen 
between the number of validation predictions within chemical 
accuracy of the DFT value for the successful transfer learning 
methods compared to direct training, with over 20\% more 
for each training dataset size. These results are visualized 
in Figure \ref{fig:full_results}. 
    
\begin{figure}[t]
\centering
\includegraphics[scale=0.55]{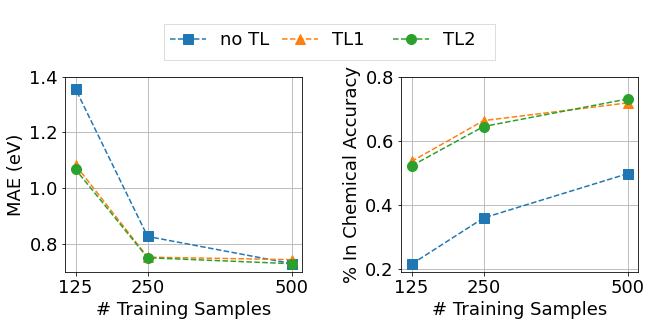}
\caption{Results of the two best transfer learning 
methods (TL1 and TL2) compared to best direct training 
method without transfer learning with the extended metal 
oxides dataset. Left: Mean absolute error (MAE) of the 
validation set evaluation 
for the models trained on each of the training set sizes. 
Right: Percent of model evaluations within chemical accuracy 
($1$ kcal/mol) of the benchmark value computed with density 
functional theory (DFT). It is seen that transfer learning 
improves significantly the number of neural net 
predictions within chemical accuracy, 
while minimizing MAE. }
\label{fig:full_results_ext}
\end{figure}

\begin{figure}
        \centering
        \hspace*{0.cm}
        \includegraphics[scale=0.51]{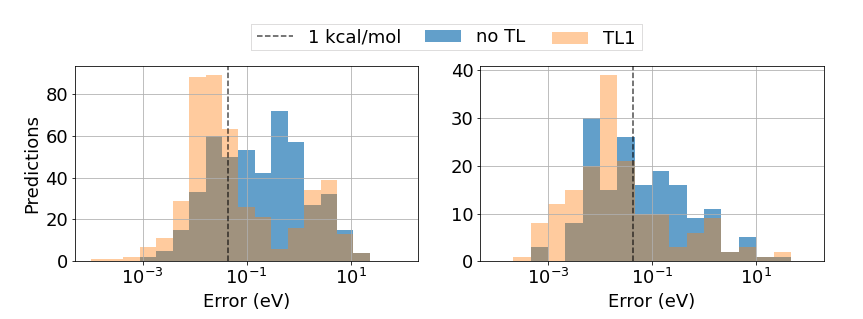}
   \caption{Error distribution in cases where TL1 and direct training had similar MAE with a significant gap in predictions within chemical accuracy. In each case, TL1 errors make up a larger portion of both the ends of the distribution, with the larger values making a significant impact to the MAE. Left: Predictions from models trained on 100 length training sets of original dataset. Right: Predictions from models trained on 500 length training set of extended 
dataset. }
        \label{fig:same_err}
    \end{figure}

Next, the first two transfer learning methods (TL1 and TL2) 
are compared to direct training on an extended version 
of the transition metal oxide dataset, which includes 
two additional transition metal elements Manganese 
and Chromium, and 146 additional data points. 
On the extended set, training is done similarly 
except with training set sizes of 500, 250, and 125 
and a validation set of 163 samples. Like the 
smaller dataset, the transfer learning methods 
both get significantly more predictions from the 
validation set within chemical accuracy of their 
DFT-calculated value than the directly 
trained model. These results are reported in Table \ref{tab:schnet_tl_results_ext} and 
Figure \ref{fig:full_results_ext}.
\begin{table}[t]
            \centering
            \begin{tabular}{c|c c c} 
                 $N$ & No TL & TL1 & TL2 \\ [0.5ex] 
                 \hline
                 125 & 1.355        &  1.083  & $\bm{1.067}$  \\ 
                 250 & 0.827        &  0.753  & $\bm{0.751}$  \\
                 500 & $\bm{0.730}$ &  0.744  & $\bm{0.730}$ 
             \end{tabular}
\caption{Validation set mean absolute error in eV (electron volts) of the best performing direct training and transfer learning methods applied to the extended metal oxides dataset.}
            \label{tab:schnet_tl_results_ext}
    \end{table}
In these two experiments, transfer learning methods 
TL1 and TL2 perform as good as or better than 
direct training in mean absolute error. However, 
even in the two cases with similar mean absolute 
error, a much larger fraction of the transfer learning 
predictions are within chemical accuracy than the 
direct model predictions. The reason for this was 
explored further, and it was found that in addition 
to the lower error, the transfer learning predictions 
also shared a higher proportion of the higher errors 
than the directly trained model, leading to similar 
mean error. This is shown for the TL1 model in 
both the 100 length training sets for the original 
dataset and the 500 length training set for the 
extended dataset in Figure \ref{fig:same_err}. 
    \begin{figure}
        \centering
        \hspace*{0cm}
        \includegraphics[scale=0.5]{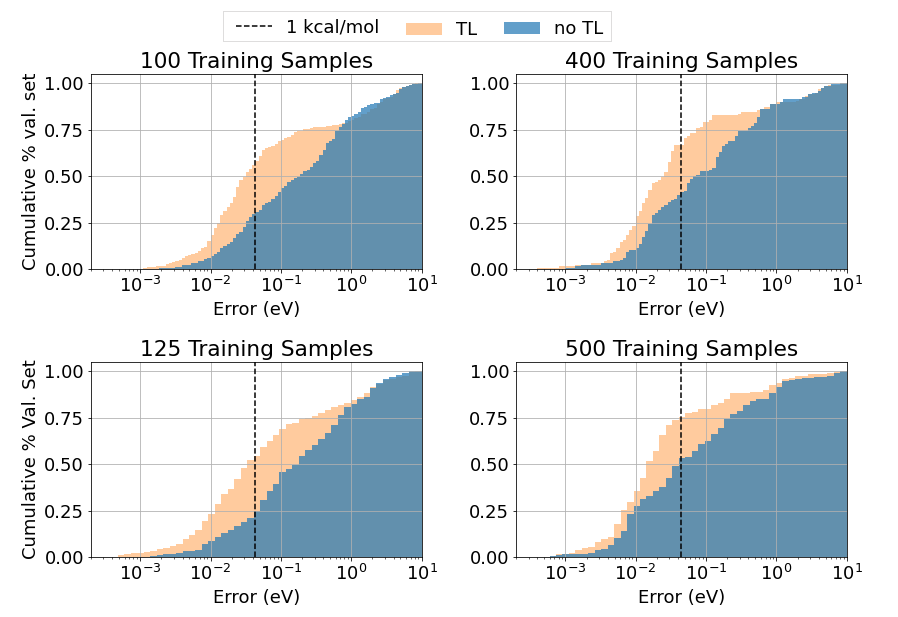}
        \caption{Top: Cumulative distribution of validation set error for direct training and TL1 methods trained with the 100 and 400 length training sets. In each case, over 20\% more of the predictions from transfer learning are within chemical accuracy (1 kcal/mol). Bottom: Cumulative distribution of validation set error for direct training and TL2 methods trained with the 125 and 500 length training sets of the extended data.}
        \label{fig:cum_err}
    \end{figure}
    
\subsection{Computational cost}
\begin{figure}[t]
            \centering
            \includegraphics[scale=0.55]{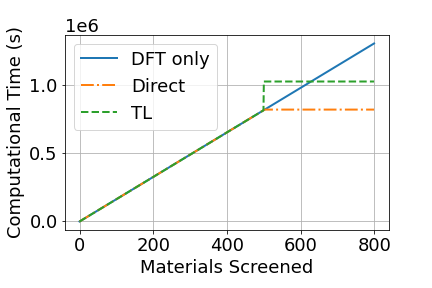}
            \caption{Projected computational time for screening properties with only DFT and neural networks. Using 500 samples for training, using either direct training or transfer learning would save 135 or 78 hours respectively in computational time over DFT in predicting properties of just 300 more materials.}
            \label{fig:comptime_comp}
\end{figure}

 As we mentioned in section \ref{sec:DFT}, 
 density functional theory calculations can be 
 computationally expensive, with the average 
 calculation from a set of 1,300 transition 
 metal oxides taking over 27 minutes per run 
 (see  Figure \ref{fig:dft_timing}).
\begin{table}[t] 
                    \centering
                    \begin{tabular}{c|cc} 
                         $N$ & No TL & TL1 \\ [0.5ex] 
                         \hline
                         125 & 7182 &  214495  \\ 
                         250 & 5008 &  210736   \\
                         500 & 3626 &  209751   
                     \end{tabular} 
                    \caption{Total computational time in seconds to train direct and transfer learning models per size of training set. Transfer learning models also take into account the training time of their source model, accounting for the large difference between methods.}
                    \label{tab:train_times}
        \end{table}
Using a neural network in lieu of DFT allows for 
accelerated predictions after the overhead cost 
of training and generation of data. With a trained 
SchNet model, predicting the free energy of a new 
transition metal oxide is in the order of milliseconds, 
while running DFT calculations takes similar time as 
the previous calculations. The overhead cost of training 
the neural network can be insignificant for larger scale 
screening of materials. 
Table \ref{tab:train_times} summarizes 
the training times for direct and transfer learning models. 
SchNet was implemented using the PyTorch machine 
learning framework in python \cite{NEURIPS2019_9015}. 
Training was done with an NVIDIA Titan RTX GPU. 
While both direct and transfer training times were similar, 
we must also consider the training of the source 
model used to initialize the transfer learning models, 
which leads to the large discrepancy between the two. 
For the largest training set, the total training time 
was 2.2 times that of the mean DFT calculation time 
for direct training, and 128.2 times for transfer learning. 
If hundreds or thousands of materials are to be screened, 
neural networks allow a large savings in computational 
time, as visualized in Figure \ref{fig:comptime_comp}.
        
To give an example of this speedup, DFT was done 
on a simple, two-atom TiO molecule using 
the Quantum Espresso DFT code on a 2014 Mac Mini 
with 1.4 GHz Intel Core i5 processor and 4GB of RAM. 
This took 87 seconds. Evaluating the same material 
with the neural network took 4.6 milliseconds on the 
same computer. While being a big speedup, this does 
not capture the more significant speedup seen with 
materials with more atoms. While DFT calculations for larger transition metal oxides took as long as 20 hours in 
the benchmark, the longest SchNet evaluation time 
was 350 milliseconds.

\section{Conclusions} 
\label{sec:conclusions}
Neural networks have become popular for 
high-throughput screening of material 
properties, as they provide significantly 
faster predictions than ab initio simulations 
based on density functional theory (DFT). 
However, in order to obtain accurate predictions
neural networks require large amounts of 
DFT training data, which can be computationally 
expensive to obtain.
To overcome this problem, in this paper we 
developed new transfer learning algorithms 
based on SchNet 
\cite{schutt_schnet_2017,schutt_schnetpack_2019}  
to repurpose trained neural network models 
on other materials using only a small amount 
of additional training data. 
%
We demonstrated that the proposed transfer learning 
algorithms can improve the mean absolute error of 
the Gibbs free energy predictions by up to $30\%$ 
compared to direct training. Furthermore, even 
in cases where the difference in the mean absolute errors 
are not significant, transfer learning increased 
the number of predictions within 
chemical accuracy by $25\%$ to $30\%$. 

\vspace{0.5cm}
\noindent 
{\bf Acknowledgements} 
This research was supported by the NSF-TRIPODS grant 81389-444168.

\bibliographystyle{plain}
\bibliography{references}

\end{document}